\documentclass{emulateapj}
\usepackage{amssymb}
\usepackage{apjfonts}
\usepackage{graphicx}
\shorttitle{The disappearance of a narrow Mg II absorption system}
\shortauthors{Chen, Qin \& Gu}

\begin{document}
\title{The disappearance of a narrow Mg II absorption system in quasar SDSS J165501.31+260517.4}

\author{Zhi-Fu Chen\altaffilmark{1,2}, Yi-Ping Qin\altaffilmark{2,4,1}, Min-Feng Gu\altaffilmark{3}}
 \altaffiltext{1}{Department of Physics and Telecommunication Engineering, Baise University, Baise, Guangxi 533000, China; zhichenfu@126.com}
 \altaffiltext{2}{Center for Astrophysics, Guangzhou University,
Guangzhou 510006, China; ypqin@126.com} \altaffiltext{3}{Key
Laboratory for Research in Galaxies and Cosmology, Shanghai
Astronomical Observatory, Chinese Academy of Sciences, 80 Nandan
Road, Shanghai 200030, China; gumf@shao.ac.cn}
\altaffiltext{4}{Physics Department, Guangxi University, Nanning
530004, P. R. China}

\begin{abstract}
In this letter, we present for the first time, the discovery of the
disappearance of a narrow Mg II $\lambda\lambda2796,2803$ absorption
system from the spectra of quasar SDSS
J165501.31+260517.4 ($z_{\rm e}=1.8671$). This absorber is located at $z_{\rm abs} =1.7877$, and has a
velocity offset of $8,423\rm ~km~s^{-1}$ with respect to the quasar.
According to the velocity offset and the line variability, this
narrow Mg II $\lambda\lambda2796,2803$ absorption system is likely
intrinsic to the quasar. Since the corresponding UV continuum
emission and the absorption lines of another narrow
Mg II $\lambda\lambda2796,2803$ absorption system at $z_{\rm abs}=1.8656$ are very stable,
we think that the disappearance of the absorption system is unlikely
to be caused by the change in ionization of absorption gas. Instead,
it likely arises from the motion of the absorption gas across the line of sight.
\end{abstract}

\section{Introduction}
It is known that absorption lines would be detected in a quasar
spectra when the quasar sight line passes through the corresponding
foreground absorbers. Absorption lines of quasars are usually split
into broad absorption lines (BALs), with the absorption troughs
being broader than $2000~\rm km~s^{-1}$ at depths $> 10\%$ below the
continuum (Weymann et al. 1991), and narrow absorption lines (NALs),
with line widths being narrower than $500~\rm km~s^{-1}$. Absorption
lines are traditionally divided into two classes according to their
relationship with the corresponding quasars: (1) the cosmologically
intervening absorption lines, which are often believed to be caused
by the cosmologically intervening galaxies lying on the quasar sight
lines (Bergeron 1986; Bond et al. 2001) and hence are physically
unrelated with the quasars; (2) the intrinsic (or associated)
absorption lines, which are physically associated with the quasars.
The intrinsic absorption lines are usually believed to be related
with quasar outflows, for which, viewing at different angles could
give rise to different line types (BALs or NALs) (Murray et al.
1995).

Variations of intrinsic absorption lines in equivalent width and/or
shape have been noted by various authors (e.g., Lundgren et al.
2007; Leighly et al. 2009; Capellupo et al. 2011; Hamann et al.
2011). It was revealed that the fractional change in equivalent
width increases with rest-frame timescale over $0.05-5$ yr. The
variation of absorption lines could arise from the changes in the
ionization state or/and in the covering factor of the absorption
gas. BALs are undoubtedly intrinsic to the corresponding quasars.
Most BAL variations are due to the changes in the covering factor of
outflow stream lines that partially block the continuum emission,
which could be caused by the motion of the outflow (e.g., Hamann
1998; Arav et al. 1999; Proga et al. 2000; Ak et al. 2012; Vivek
2012). While the variation of BALs would be common, extreme events
such as the disappearance and the emergence of absorption troughs
from the spectra are rare (which might consume many years of
observation). So far, only a small number of these extreme events of
BALs have been reported (e.g., Hall et al. 2011; Vivek et al. 2012;
Filiz et al. 2012). Variations in the ionization state of absorption
gas are also common. In a recent study, the coordinated line
variations of five narrow intrinsic absorption systems imprinted in
one quasar spectra could be best explained by global changes in the
ionization of absorption gas due to the changes in the quasar's
ionizing emission (Hamann et al. 2011). However, no disappearance
and emergence of narrow absorption lines have ever been reported.

Quasars are capable of driving outflows with speeds up to
$30,000~\rm km~s^{-1}$ (Ganguly et al. 2007). Therefore, the narrow
intrinsic absorption lines can occur at large velocity separations
from the quasars (e.g., Misawa et al. 2007; Ganguly \& Brotherton
2008; Hamann et al. 2011). In addition, narrow absorption lines
imprinted in the quasar spectra are very common for intervening
absorption lines. Thus, in many cases, confirming a narrow
absorption line to be intrinsic to the quasar is very difficult (see
Ganguly \& Brotherton 2008 for a review).

In this letter, we analyze the spectra of quasar SDSS J165501.31+260517.4
and report our discovery of the disappearance of a narrow
Mg II $\lambda\lambda2796,2803$ absorption system. Throughout this
paper we use cosmological parameters $\Omega_\Lambda=0.7$,
$\Omega_M=0.3$ and $H_{0}= 70 ~\rm km~ s^{-1}~ Mpc^{-1}$.

\section{Analysis}
There are about 8000 quasars included in both the SDSS-III (Sloan
Digital Sky Survey-III) quasar catalog (Eisenstein et al. 2011; Ross
et al. 2012) and the SDSS-II quasar catalog (York et al. 2000;
Schneider et al. 2010), meaning that these quasars have been
observed twice by the Sloan Digital Sky Survey. This large sample
provides us a chance to search possible disappearance of absorption
systems. Among these quasars, 33 were detected to possess obvious Mg
II $\lambda\lambda2796,2803$ associated absorption systems by Shen
\& M\'enard (2012) with the data of the SDSS-II quasar catalog. To
investigate the variability of the narrow absorption systems, we
analyzed the spectra of all these 33 quasars available in both
SDSS-I/II and SDSS-III. We fitted each spectrum with a combination
of cubic splines for pseudo-continuum, and Gaussians for line
features. The Mg II $\lambda\lambda2796,2803$ absorption systems
(including all possible associated and intervening absorption
systems) were searched in the pseudo-continuum normalized spectra.

We found that all Mg II $\lambda\lambda2796,2803$ associated
absorption systems in the spectra of SDSS-I/II, noticed by Shen \&
M\'enard (2012), are also detected in those of SDSS-III. However, in
quasar SDSS J165501.31+260517.4 ($z_{\rm e}=1.8671$, Hewett \& Wild
2010), we accidently found that the narrow Mg II
$\lambda\lambda2796,2803$ absorption system at $z_{\rm abs}
=1.7877$, observed on 12 April 2005 (SDSS-I/II), disappears from the
SDSS-III spectra which was observed on 13 April 2011. We show the
spectra of this quasar together with the corresponding
pseudo-continua in Fig. 1, and present the pseudo-continuum
normalized spectra in Fig. 2. The detection and disappearance of Mg
II $\lambda\lambda2796,2803$ absorption system can be clearly seen
in Fig. 2 by comparing SDSS-I/II and SDSS-III spectra. Each
absorption line of the detected Mg II $\lambda\lambda2796,2803$
absorption system in the SDSS-I/II spectra was fitted with a
Gaussian component, from which we measured the rest-frame equivalent
width ($W_{\rm r}$). We estimate the uncertainty of the detected
absorption lines via
\begin{equation}
(1+z)\sigma_w=\frac{\sqrt{\sum_i
P^2(\lambda_i-\lambda_0)\sigma^2_{f_i}}}{\sum_i
P^2(\lambda_i-\lambda_0)}\Delta\lambda
\end{equation}
here $P(\lambda_i-\lambda_0),~\lambda_i,~and~\sigma_{f_i}$
represent the line profile centered at $\lambda_0$, the wavelength,
and the normalized flux uncertainty as a function of pixel (Nestor
et al. 2005; Quider et al. 2011). The sum is performed over an
integer number of pixels that cover at least $\pm 3$ characteristic
Gaussian widths. The measurements of the detected absorption lines
are presented in Table 1.

\begin{figure*}
\centering
\includegraphics[width=16 cm,height=6 cm]{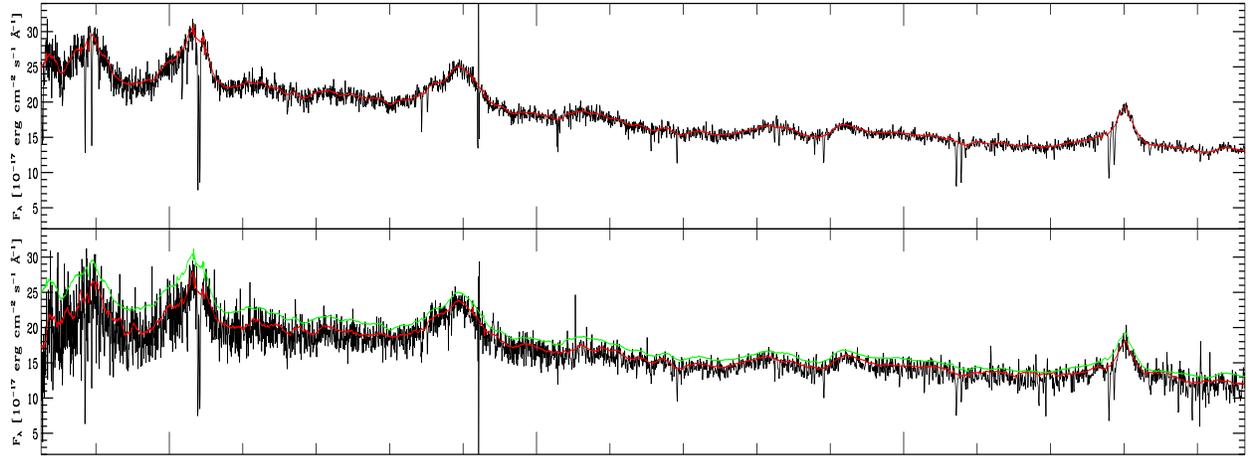}
\caption{The spectra of quasar SDSS J165501.31+260517.4, observed by
SDSS-I/II (the lower panel) and SDSS-III (the upper panel)
respectively. The red solid lines represent the pseudo-continua, and
the solid green line presented in the lower panel is the
pseudo-continuum shown in the upper panel.}
\end{figure*}

\begin{figure*}
\includegraphics[width=16 cm,height=4.5 cm]{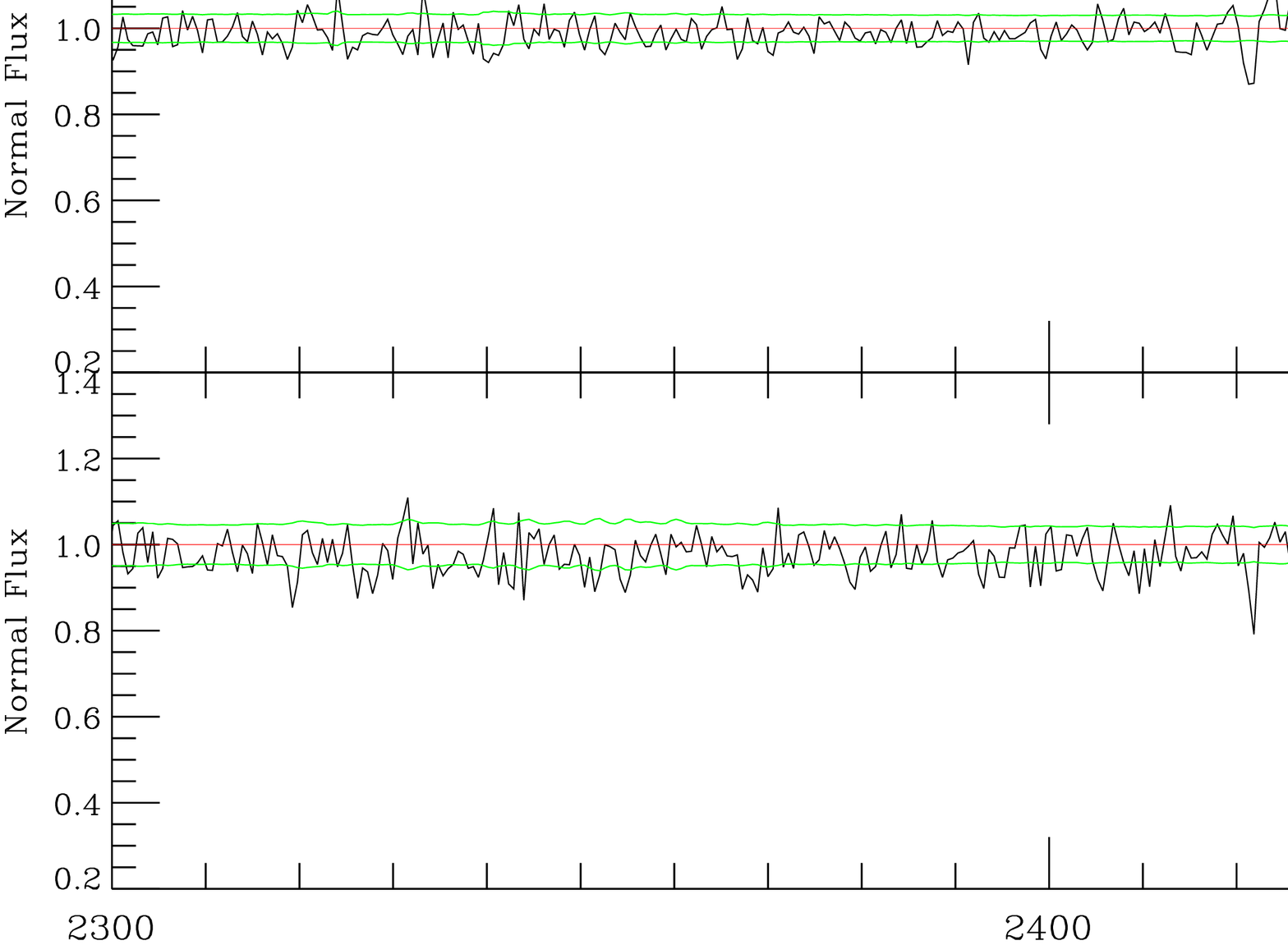}
\includegraphics[width=16 cm,height=4.5 cm]{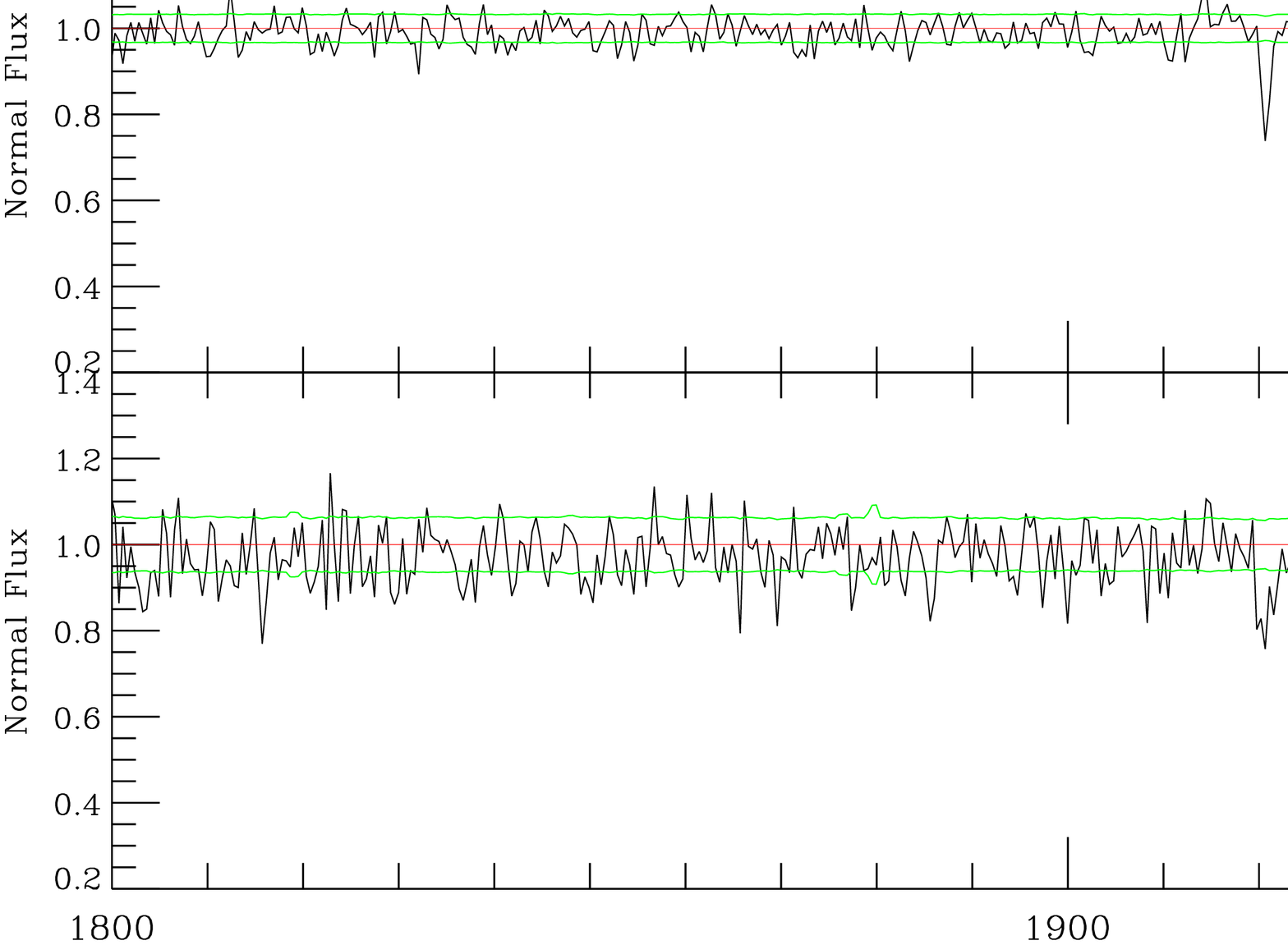}
\includegraphics[width=16 cm,height=4.5 cm]{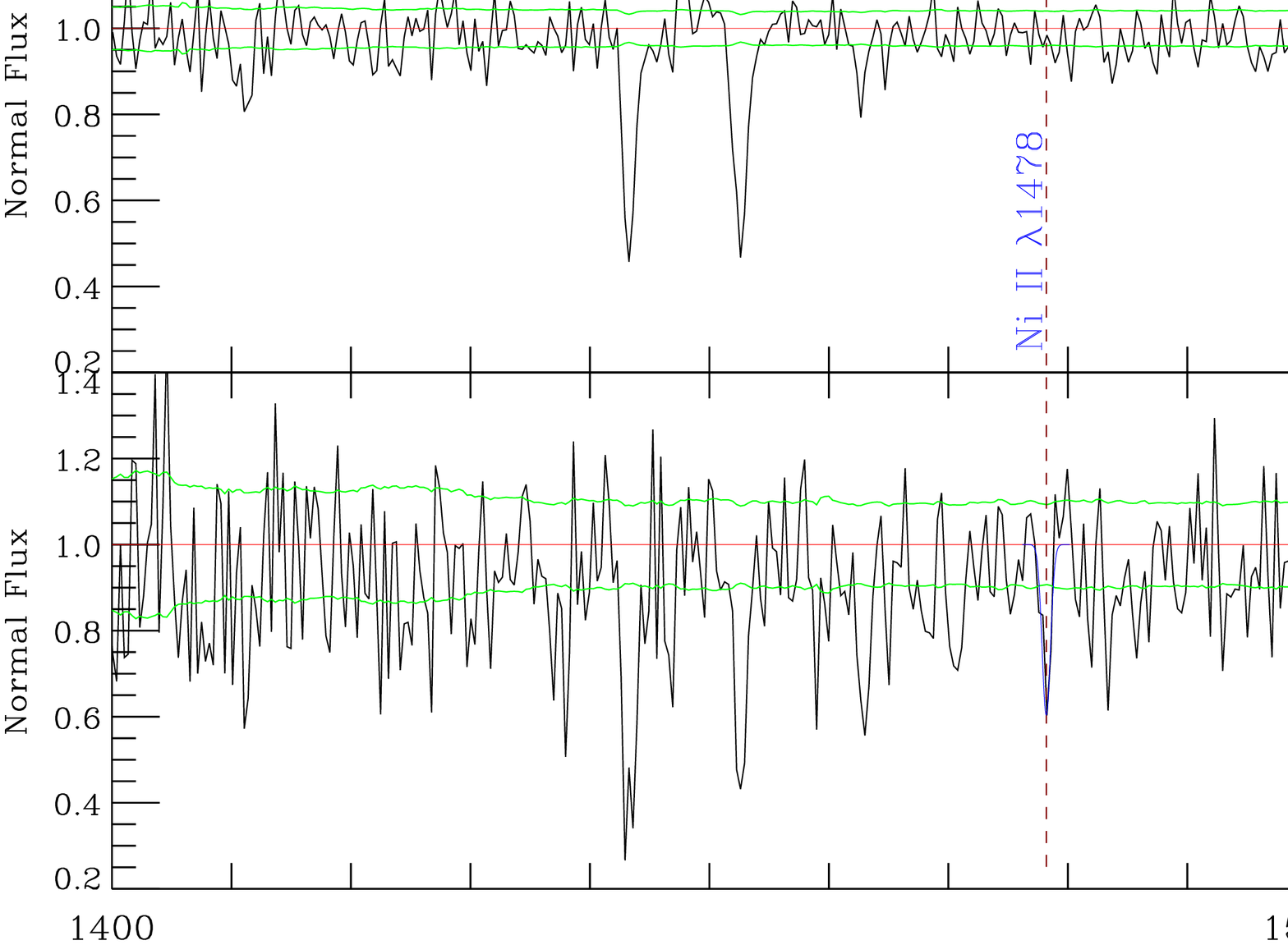}
\caption{The pseudo-continuum normalized spectra of quasar SDSS
J165501.31+260517.4, observed by SDSS-I/II (the lower panel) and
SDSS-III (the upper panel) respectively. The solid green lines
represent the flux uncertainty levels which have been normalized by
the corresponding pseudo-continuum, the solid blue curves represent
the Gaussian fitting, and the dashed dark red lines show the
positions of absorption lines. A Mg II $\lambda\lambda2796,2803$
absorption system with $z_{\rm abs}=1.7877$ is shown in the lower panel
(see the solid blue curves).}
\end{figure*}

\begin{table}
\caption{\scriptsize Parameters of disappearing absorption lines}
\tabcolsep 2.5mm
 \begin{tabular}{cccccc}
 \hline\hline\noalign{\smallskip}
Species&$W_r$(\AA)&$W_r$(\AA)$^{a}$&\\
 \hline
$Mg~II\lambda2803$&0.33$\pm$0.10&0.05\\
$Mg~II\lambda2796$&0.51$\pm$0.11&0.05\\
$Fe~II \lambda2600$&0.28$\pm$0.16&0.06\\
$N~III\lambda1749$&0.18$\pm$0.06&0.06\\
$Ni~II\lambda1741$&0.12$\pm$0.05&0.06\\
$Fe~II\lambda1563$&0.16$\pm$0.08&0.06\\
$Si~II\lambda1527$&0.15$\pm$0.10&0.06\\
$Fe~II\lambda1504$&0.50$\pm$0.20&0.06\\
$Ni~II\lambda1478$&0.35$\pm$0.10&0.06\\
\hline\noalign{\smallskip}
\end{tabular}
\\
$^{a}$The equivalent width limits estimated from the corresponding
SDSS-III spectrum are also calculated by Equation (1).
\end{table}

\section{Discussion and conclusions}
Although the disappearance of BALs was reported by various authors
previously (e.g., Hall et al. 2011; Vivek et al. 2012; Filiz Ak et
al. 2012), so far, the disappearance of narrow absorption lines has
not been detected yet. Here, we show the first discovery of the
disappearance of a narrow Mg II $\lambda\lambda2796,2803$ absorption
system in quasar SDSS J165501.31+260517.4. The velocity offset of
the corresponding absorber ($z_{\rm abs}=1.7877$) with respect to
the quasar ($z_{\rm e}=1.8671$) is $8,423\rm ~km~s^{-1}$. The line
variability together with the velocity offset suggest that this
narrow Mg II $\lambda\lambda2796,2803$ absorption system is likely
to be intrinsic to the corresponding quasar.

Two mechanisms are likely responsible for the disappearance of the
narrow Mg II $\lambda\lambda2796,2803$ absorption system: the motion
of the absorbing gas across the line of sight, and the fluctuation
of quasar ionized radiation which could give rise to the change in
the ionization of absorption gas. To check whether the fluctuation
of quasar ionized radiation is the origin of the disappearance, let
us study the line variability of other narrow Mg II
$\lambda\lambda2796,2803$ absorption systems detected in the same
spectra. In this quasar, we have detected another narrow Mg II
$\lambda\lambda2796,2803$ absorption system at $z_{\rm abs}=1.8656$
identified in both the SDSS-I/II and SDSS-III spectra, observed on
12 April 2005 and 13 April 2011, respectively. The velocity offset
of this absorber with respect to the quasar is $1,562\rm ~km~s^{-1}$
(see Fig. 2). We found that this absorption doublet is very stable
(within $1\sigma$ error) during two epochs, with $W_{\rm
r}(\lambda2796) = 0.93 \pm 0.09\AA$~and $W_{\rm r}(\lambda2803) =
0.72 \pm 0.09 \AA$ from the SDSS-III spectrum, and $W_{\rm
r}(\lambda2796) = 0.99 \pm 0.17\AA$ and $W_{\rm r}(\lambda2803) =
0.66 \pm 0.15\AA$ from the SDSS-I/II spectrum. If the disappearance
of the $z_{\rm abs}=1.7877$ narrow Mg II $\lambda\lambda2796,2803$
absorption system is caused by the fluctuation of quasar ionized
radiation, it is unlikely that the $z_{\rm abs}=1.8656$ narrow Mg II
$\lambda\lambda2796,2803$ absorption system can be so stable within
the same period. In addition, we found that the near-UV continuum
fluxes of this quasar are stable at  two epochs. Therefore, the
fluctuation of the quasar ionized radiation is unlikely the origin
of the disappearance.

\begin{figure}
\centering
\includegraphics[width=5 cm,height=4.5 cm]{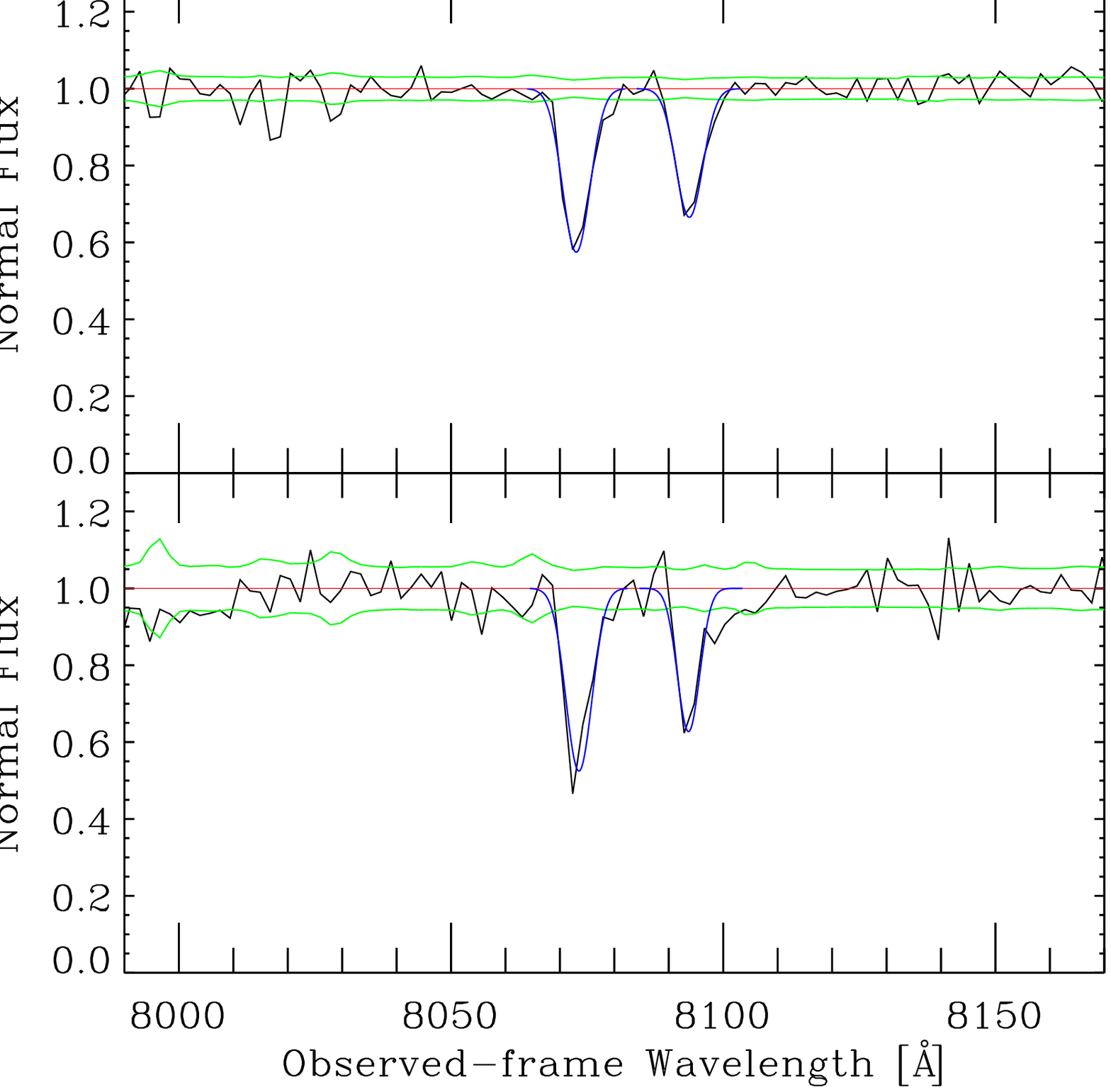}
\vspace{2.5ex}
\caption{The Mg II $\lambda\lambda2796,2803$ absorption system at
$z_{\rm abs}=1.8656$ imprinted in the spectra of quasar SDSS
J165501.31+260517.4. The lower panel is from SDSS-I/II and the upper
panel is from SDSS-III. The lines are the same as in Fig. 2.}
\end{figure}

Another mechanism likely accounting for the disappearance is the
movement of the absorption gas. It is possible that a clumpy gas
partially covers the background emission source, and its motion
might cause the disappearance of the narrow Mg II
$\lambda\lambda2796,2803$ absorption system if the transverse motion
of the gas makes it passing through our line of sight, especially
when the absorber has a sharp edge and the sightline is close to its
edge. Considering the case of a geometric thin and optical thick
accretion disk, most of the UV continuum radiation is expected to
originate from the inner accretion disk, whose size is in the order
of $D_{\rm cont} \sim 5R_{\rm S}=10GM_{\rm BH}/c^2$ (Wise et al.
2004; Misawa et al. 2005). Here, let us take the virial black hole
mass $M_{\rm BH}=10^{9.2}~M_\odot$, estimated based on the Mg II
broad emission line (Shen et al. 2011), as the mass of the black
hole. That gives rise to $D_{cont}=1.5\times 10^{10.2}~km$. The
disappearance of the narrow Mg II $\lambda\lambda2796,2803$
absorption system occurred within a period of 765 days in the quasar
rest-frame. We assume a face-on accretion disk and that the
transverse size of the absorption gas is as large as the UV
continuum emitting region. Taking 765 days as an upper limit of the
transverse motion time of the absorption gas, one can estimate the
value of the transverse velocity perpendicular to the sightline,
which is about $360~\rm km~s^{-1}$. Adopting the virial black hole
mass and assuming that the shift velocity of the gas does not exceed
the escape velocity, we can constrain the location of the absorber
relative to the central region: the gas locates at a radius of $r
\sim 0.1~pc$ (see, e.g., Misawa et al. 2005). According to the
empirical relationship between the radius and luminosity, the radius
of the broad emission line region (BLR) is $R_{\rm BLR} \approx
~0.08~pc$, which was estimated from the continuum luminosity at 1350
\AA~ (Kaspi et al. 2007), where the continuum luminosity at 1350
\AA~ is directly taken from Shen et al. (2011). In this way, the
absorber would locate at the vicinity of the broad emission line
region.

\section*{Acknowledgements} We thank the anonymous referee for helpful comments
and suggestions. This work was supported by the 973 Program (No.
2009CB824800), the National Natural Science Foundation of China (No.
11073007, and 11073039), the Guangzhou technological project (No.
11C62010685), and Guangxi Natural Science Foundation
(2012jjAA10090).


\begin{thebibliography}{99}
\bibitem[\protect\citeauthoryear{}{}]{}Arav N., Becker R. H., Laurent-Muehleisen S. A. et al., 1999, ApJ, 524, 566
\bibitem[\protect\citeauthoryear{}{}]{}Ak F. N., Brandt W. N., Hall P. B. et al., 2012, ApJ, 757, 114
\bibitem[\protect\citeauthoryear{}{}]{}Bergeron J., 1986, A\&A, 155, L8
\bibitem[\protect\citeauthoryear{}{}]{}Bond N. A., Churchill C.W., Charlton J. C., Vogt S. S., 2001, ApJ, 562, 641
\bibitem[\protect\citeauthoryear{}{}]{}Capellupo D. M., Hamann F., Shields J. C. et al., 2011, MNRAS, 413, 908
\bibitem[\protect\citeauthoryear{}{}]{}Eisenstein D. J., Weinberg D. H., Agol E. et al., 2011, AJ, 142, 72
\bibitem[\protect\citeauthoryear{}{}]{}Filiz A. N., Brandt W. N., Hall P. B. et al., 2012, ApJ, 757, 114
\bibitem[\protect\citeauthoryear{}{}]{}Ganguly R., Scoggins B., Cales S., Brotherton M. S., \& Vestergaard M. 2007, ApJ, 665, 990
\bibitem[\protect\citeauthoryear{}{}]{}Ganguly R., Brotherton M. S., 2008, ApJ, 672, 102
\bibitem[\protect\citeauthoryear{}{}]{}Hall P. B., Anosov K., White R. L. et al., 2011, MNRAS, 411, 2653
\bibitem[\protect\citeauthoryear{}{}]{}Hamann, F. 1998, ApJ, 500, 798
\bibitem[\protect\citeauthoryear{}{}]{}Hewett P. C. \& Wild V., 2010, MNRAS, 405, 2302
\bibitem[\protect\citeauthoryear{}{}]{}Hamann F., Kanekar N., Prochaska J. X. et al., 2011, MNRAS, 410, 1957
\bibitem[\protect\citeauthoryear{}{}]{}Kaspi S., Brandt W. N., Maoz D., Netzer H. et al., 2007, ApJ, 659, 997
\bibitem[\protect\citeauthoryear{}{}]{}Leighly K. M., Hamann F., Casebeer D. A., Grupe D., 2009, ApJ, 701, 176
\bibitem[\protect\citeauthoryear{}{}]{}Lundgren B. F., Wilhite B. C., Brunner R. J. et al., 2007, ApJ, 656, 73
\bibitem[\protect\citeauthoryear{}{}]{}Murray N., Chiang J., Grossman S. A., Voit G. M., 1995, ApJ, 451, 498
\bibitem[\protect\citeauthoryear{}{}]{}Misawa T., Eracleous M. D., Charlton J. C., Tajitsu A., 2005, ApJ, 629, 115
\bibitem[\protect\citeauthoryear{}{}]{}Misawa T., Charlton J. C., Eracleous M. D. et al., 2007, ApJS, 171, 1
\bibitem[\protect\citeauthoryear{}{}]{}Nestor D. B., Turnshek D. A., Rao S. M., 2005, ApJ, 628, 637
\bibitem[\protect\citeauthoryear{}{}]{}Proga D., Kallman T. R., Drew J. E., Hartley L. E., 2002, ApJ, 572, 382
\bibitem[\protect\citeauthoryear{}{}]{}Quider A. M., Nestor D. B., Turnshek D. A. et al., 2011, AJ, 141, 137
\bibitem[\protect\citeauthoryear{}{}]{}Ross N. P., Myers A. D., Sheldon E. S. et al., ApJS, 2012, 199, 3
\bibitem[\protect\citeauthoryear{}{}]{}Schneider D. P., Richards G. T., Hall P. B. et al., 2010, AJ, 139, 2360
\bibitem[\protect\citeauthoryear{}{}]{}Shen Y., Richards G., Strauss M. A.,  et al., 2011, ApJS, 194, 45
\bibitem[\protect\citeauthoryear{}{}]{}Shen Y. \& M\'enard B., 2012, ApJ, 748, 131
\bibitem[\protect\citeauthoryear{}{}]{}Vivek M., Srianand R., Mahabal A., \& Kuriakose V. C., 2012, MNRAS, 421, L107
\bibitem[\protect\citeauthoryear{}{}]{}Weymann R. J., Morris S. L., Foltz C. B., Hewett P. C., 1991, ApJ, 373, 23
\bibitem[\protect\citeauthoryear{}{}]{}Wise J. H., Eracleous M., Charlton J. C., Ganguly R., 2004, ApJ, 613, 129
\bibitem[\protect\citeauthoryear{}{}]{}York D. G., Adelman J., Anderson J. E. Jr. et al., 2000, AJ, 120, 1579


\end{thebibliography}
\end{document}